\documentclass[journal,12pt,onecolumn]{IEEEtran}

\usepackage{graphics}
\usepackage{rotating}
\usepackage{amsfonts}
\usepackage{amsmath}
\usepackage{subfigure}
\usepackage{comment}

\begin{document}
\title{\Large\bf
On ``A Novel Maximum Likelihood Decoding Algorithm for\\
Orthogonal Space-Time Block Codes''\\[5mm]}
\author{
\normalsize
Ender Ayanoglu\\
Center for Pervasive Communications and Computing\\
Department of Electrical Engineering and Computer Science\\
University of California, Irvine}

\maketitle

\begin{abstract}
  The computational complexity of the Maximum Likelihood
  decoding algorithm in \cite{aa}, \cite{aa2}
  for orthogonal space-time block codes is smaller than specified.
\end{abstract}

\section{Introduction}
In \cite{aa},\cite{aa2}, the decoding of an Orthogonal Space-Time
Block Code (OSTBC) in a Multi-Input Multi-Output (MIMO) system with
$N$ transmit and $M$ receive antennas, and an interval of $T$ symbols
during which the channel is constant, is considered. The received
signal is given by 
\begin{equation}
Y={\cal G}_N H+V
\label{eq-1}
\end{equation}
where $Y=[y_t^j]_{T\times M}$ is the received signal matrix of size
$T\times M$ and whose entry $y_t^j$ is the signal received at antenna
$j$ at time $t$, $t=1,2,\ldots,T$, $j=1,2\ldots,M$; $V=[v_t^j]_{T\times
  M}$ is the noise matrix, and ${\cal G}_N=[g_t^j]_{T\times N}$ is the
transmitted signal matrix whose entry $g_t^i$ is the signal
transmitted at antenna $i$ at time $t$, $t=1,2,\ldots,N$. The matrix
$H=[h_{i,j}]_{N\times M}$ is the channel coefficient matrix of size
$N\times M$ whose entry $h_{i,j}$ is the channel coefficient from
transmit antenna $i$ to receive antenna $j$. The entries of the
matrices $H$ and $V$ are independent, zero-mean, and circularly
symmetric complex Gaussian random variables.

The  real-valued representation of (\ref{eq-1}) is obtained by
first arranging the matrices $Y$, $H$, and $V$, each in one column
vector by stacking their columns one after the other as 
\begin{equation}
\left[\begin{array}{c}y_1^1\\ \vdots \\ y_T^M \end{array} \right] =
\check {\cal G}_N 
\left[\begin{array}{c}h_{1,1}\\ \vdots \\ h_{N,M} \end{array} \right]
+
\left[\begin{array}{c}v_1^1\\ \vdots \\ v_T^M \end{array} \right] 
\label{eq-2}
\end{equation}
where $\check{\cal G}_N = I_M \otimes {\cal G}_N$, with $I_M$ denoting
the identity matrix of size $M$ and $\otimes$ denoting the Kronecker
matrix multiplication, and then decomposing the $MT$-dimensional
complex problem defined by (\ref{eq-2}) to a $2MT$-dimensional
real-valued problem by applying the real-valued lattice representation
defined in \cite{aa3} to obtain 
\begin{equation}
\check y = \check H x + \check v
\label{eq-3}
\end{equation}
or equivalently
\begin{equation}
\left[\begin{array}{c}{\rm Re}(y_1^1)\\{\rm Im}(y_1^1) \\ \vdots \\
 {\rm Re}(y_T^M)\\ {\rm Im}(y_T^M) \end{array} \right] =
\check H
\left[\begin{array}{c}{\rm Re}(s_1)\\{\rm Im}(s_1) \\ \vdots \\
 {\rm Re}(s_K)\\ {\rm Im}(s_K) \end{array} \right] 
+
\left[\begin{array}{c}{\rm Re}(v_1^1)\\{\rm Im}(v_1^1) \\ \vdots \\
 {\rm Re}(v_T^M)\\ {\rm Im}(v_T^M) \end{array} \right]  .
\label{eq-4}
\end{equation}

The real-valued fading coefficients of $\check H$ are defined using
the complex fading coefficients $h_{i,j}$ from transmit antenna $i$ to
receive antenna $j$ as $h_{2l-1+2(j-1)N}={\rm Re}(h_{l,j})$ and 
$h_{2l+2(j-1)N}={\rm Im}(h_{l,j})$ for $l=1,2,\ldots,N$ and
    $j=1,2,\ldots,M$. Since ${\cal G}_N$ is an orthogonal matrix and
    due to the real-valued representation of the system using
    (\ref{eq-4}), it can be observed that
\begin{itemize}
\item All columns of $\check H = [\check h_1 \check h_2 \ldots \check
  h_{2K}]$ where $\check h_i$ is the $i$th column of $\check H$, are
  orthogonal to each other, or equivalently
\begin{equation}
\check h_i^T\check h_j = 0 \hspace{1cm}
i, j =1, 2, \ldots, K, i\neq j
\label{eq-5}
\end{equation}
\item The inner product of every column in $\check H$ with itself
  is equal to a constant, i.e.,  
\begin{equation}
\check h_i^T \check h_i =  \check h_j^T \check h_j \hspace{1cm} 
i, j =1, 2, \ldots, K.
\label{eq-6}
\end{equation}
\end{itemize}

\section{Decoding}
Let 
\begin{equation}
\sigma = \check h_1^T\check h_1 . 
\label{eq-7}
\end{equation}
Note $\sigma = \check
h_i^T\check h_i$, $i=1, 2, \ldots, 2K$. Due to the orthogonality
property in (\ref{eq-5})-(\ref{eq-6}), we have
\begin{equation}
\check H^T \check H = \sigma I_{2K}.
\label{eq-8}
\end{equation}
Let's represent (\ref{eq-4}) as 
\begin{equation}
\check y=\check H x + \check v .
\label{eq-9}
\end{equation}
By multiplying this expression by $\check H^T$ on the left, we have 
\begin{eqnarray}
\bar y & = & \check H^T\check y \label{eq-10}\\
       & = & \sigma x + \bar v 
\label{eq-11}
\end{eqnarray}
where $\bar v$ is zero-mean, and due to (\ref{eq-8}) has independent
and identically distributed Gaussian members. The Maximum
Likelihood solution is found by minimizing 
\begin{equation}
\left\|\left[
\begin{array}{c}\bar y_1\\ \bar y_2 \\ \vdots \\ \bar y_{2K}\end{array}
\right] -
\sigma
\left[
\begin{array}{c}\bar x_1\\ \bar x_2 \\ \vdots \\ \bar x_{2K}\end{array}
\right] 
\right\|_2^2
\label{eq-12}
\end{equation}
over all combinations of $x\in \Omega^{2K}$. This can be further
simplified as 
\begin{equation}
\hat x_i = \arg \min_{x_i\in \Omega} |\bar y_i - \sigma x_i |^2
\label{eq-13}
\end{equation}
for $i=1,2,\ldots, 2K$. Then, the decoded message is
\begin{equation}
\hat x = (\hat x_1, \hat x_2,\ldots,\hat x_{2K})^T.
\label{eq-14}
\end{equation}
\section{Computational Complexity}
The decoding operation consists of the multiplication 
$\check H^T\check y$, calculation of $\sigma = \check h_1^T\check
h_1$, the multiplications $\sigma x_i$, and performing
(\ref{eq-13}). With a slight change, we will consider the calculation
of $\sigma^{-1}$ and the multiplications 
\begin{equation}
z_i = \sigma^{-1}\bar y_i \hspace{1cm}i=1,2,\ldots,2K. 
\label{eq-15}
\end{equation}
Then
\begin{equation}
\hat x_i =\arg \min_{x_i\in \Omega}|z_i - x_i |^2
\label{eq-16}
\end{equation}
for $i=1,2,\ldots,2K$, which is a standard quantization operation in
conventional Quadrature Amplitude Modulation. We
will compute the decoding complexity up to this quantization
operation. Note $\check H$ is a $2MT\times 2K$ matrix, $\check h_1$ is
a $2MT$-dimensional vector, and we will assume the complexity of real
division as equivalent to 4 real multiplications as in
\cite{aa},\cite{aa2}. The multiplication $\check H^T\check y$ takes
$2MT\cdot 2K$, calculation of $\sigma$ takes $2MT$, its
inverse takes 4, and $\sigma^{-1}\bar y$ takes $2K$ real
multiplications. Similarly, the multiplication $\check H^T\check y$
takes $2K \cdot (2MT-1)$, and calculation of $\sigma$ takes $2MT-1$
real additions. Letting $R_M$ and $R_A$ be the number of real
multiplications and the number of real additions, the complexity of
decoding the transmitted complex signal $(s_1,s_2,\ldots,s_K)$ with
the technique described in (\ref{eq-7}), (\ref{eq-10}), 
and (\ref{eq-15}) is
\begin{equation}
{\cal C}_{\rm PR}=(4KMT+2MT+2K+4)R_M,(4KMT+2MT-2K-1)R_A
\label{eq-17}
\end{equation}
which is smaller than the complexity specified in \cite{aa},\cite{aa2}
and does not depend on the constellation size $L$. However, as will be
seen in the examples, the matrix $\check H$ can include values
identical to 0, or multiplications by a scalar, which result in
deviations from (\ref{eq-17}). Also, in (\ref{eq-56}), we will provide a 
slightly smaller figure for this complexity. In what follows, 
we will calculate the exact complexity values for four examples.

Due to the orthogonality property (\ref{eq-8}) of $\check H$, the QR
decomposition of $\check H$ is 
\begin{equation}
Q=\frac{1}{\sqrt{\sigma}}\check H\hspace{1cm}R=
\sqrt{\sigma}I_{2K}
\label{eq-18}
\end{equation}
and therefore does not need to be computed explicitly. The procedure
described above is equivalent and has lower computational complexity.
\section{Comparison with a Conventional Technique}
We will now compare the technique in (\ref{eq-7}), (\ref{eq-10}), and
(\ref{eq-15})  
with one from the literature. In
\cite{ls}, it has been shown that 
\begin{equation}
\| Y - {\cal G}_NH\|^2 = \| H\|^2 \sum_{k=1}^K |s_k-\hat s_k|^2 + 
{\rm constant,}
\label{eq-19}
\end{equation}
where 
\begin{equation}
\hat s_k = \frac{1}{\| H\|^2}[{\rm Re}\{{\rm Tr}(H^HA_k^HY)\}-\hat\imath\cdot
  {\rm Im}\{{\rm Tr}(H^HB_k^HY)\}]
\label{eq-20}
\end{equation}
and where $A_k$ and $B_k$ are the matrices in
the linear 
representation of ${\cal G}_N$ in terms of $\bar s_k={\rm Re} [s_k]$
and $\tilde s_k={\rm Im} [s_k]$ for $k=1,2,\ldots,K$ as \cite{ls} 
\begin{equation}
{\cal G}_N=\sum_{k=1}^K{\bar s_k}A_k+\hat\imath {\tilde s_k}B_k 
= \sum_{k=1}^K s_k{\check A}_k+s_k^*{\check B}_k,
\label{eq-21}
\end{equation}
$\hat\imath = \sqrt{-1}$, $A_k=\check A_k+\check B_k$, and $B_k=\check
A_k-\check B_k$. 
Once $\{\hat s_k\}_{k=1}^K$ are calculated, the decoding problem can
be solved by
\begin{equation}
\min_{s_k\in \Omega^2}|s_k-\hat s_k|^2
\label{eq-22}
\end{equation}
once for each $k=1,2,\ldots,K$. Similarly to (\ref{eq-16}), this is a
standard quantization problem in Quadrature Amplitude Modulation and
we will calculate the computational complexity of this approach up to
this point. 

We will carry out the computational complexity analysis of the
technique in (\ref{eq-7}), (\ref{eq-10}), and (\ref{eq-15})
against the complexity of the technique in 
(\ref{eq-20}) for four examples, including those in
\cite{aa}, \cite{aa2}.  

{\em Example 1:\/} Consider the Alamouti OSTBC with $N=K=T=2$ and $M=1$
where 
\begin{equation}
{\cal G}_2 = \left [\begin{array}{cc}s_1&s_2\\ -s_2^*&s_1^*\\
\end{array}\right ] .
\label{eq-23}
\end{equation}
The received signal is given by
\begin{equation}
\left[\begin{array}{c}y_1\\ y_2\\\end{array}\right]=
\left [\begin{array}{cc}s_1&s_2\\ -s_2^*&s_1^*\\\end{array}\right ] 
\left[\begin{array}{c}h_{1,1}\\ h_{2,1}\\\end{array}\right] +
\left[\begin{array}{c}v_1\\ v_2\\\end{array}\right]
\label{eq-24}
\end{equation}
Representing (\ref{eq-24}) in the real domain, we have
\begin{equation}
\left[\begin{array}{c}{\rm Re}(y_1)\\{\rm Im}(y_1)\\{\rm Re}(y_2)\\{\rm Im}(y_2)\\\end{array}
\right]=\check H
\left[\begin{array}{c}x_1\\x_2\\x_3\\x_4\\\end{array}
\right] + 
\left[\begin{array}{c}{\rm Re}(v_1)\\{\rm Im}(v_1)\\{\rm Re}(v_2)\\{\rm Im}(v_2)\\\end{array}
\right]
\label{eq-25}
\end{equation}
where $x_1={\rm Re}(s_1)$, $x_2={\rm Im}(s_1)$, $x_3={\rm Re}(s_2)$, $x_4={\rm Im}(s_2)$
and
\begin{equation}
\check H = \left[\begin{array}{cccc}
h_1&-h_2&h_3&-h_4\\
h_2&h_1&h_4&h_3\\
h_3&h_4&-h_1&-h_2\\
h_4&-h_3&-h_2&h_1\\
\end{array}\right] .
\label{eq-26}
\end{equation}
Note that the matrix $\check H$ is orthogonal and all of its columns
have the same squared norm. One needs 16 real multiplications 
to calculate $\bar y=\check H^T\check y$, 4 real multiplications to calculate
$\sigma=\check h_1^T\check h_1$, 4 real multiplications to calculate
$\sigma^{-1}$, and 4 real multiplications to calculate $\sigma^{-1}\bar
y$. There are $3\cdot 4=12$ real
additions to calculate $\check H^T\check y$ and 3 real additions to
calculate $\sigma$. As a result, with this approach, decoding takes a total
of 28 real multiplications and 15 real additions. 

For the method in (\ref{eq-20}) above, the products
$H^HA_1^H$, $H^HB_1^H$, $H^HA_2^H$, $H^HB_2^H$ are
\begin{equation}
\begin{array}{cc}
H^HA_1^H = [\begin{array}{cc} h_{1,1}^* &  h_{2,1}^*\end{array}]\hspace{5mm} &
H^HB_1^H = [\begin{array}{cc} h_{1,1}^* & -h_{2,1}^*\end{array}] \\
H^HA_2^H = [\begin{array}{cc} h_{2,1}^* & -h_{1,1}^*\end{array}]\hspace{5mm} &
H^HB_2^H = [\begin{array}{cc} h_{2,1}^* &  h_{1,1}^*\end{array}] \\
\end{array}
\label{eq-27}
\end{equation}
which will be multiplied by $Y=(y_1, y_2)^T$ where $h_{1,1}$,
$h_{2,1}$, $y_1$, $y_2$ are all complex. It can be observed from
(\ref{eq-20}) and (\ref{eq-27}) that one needs all products $h_{i,1}^*y_j$,
$i,j=1,2$. Therefore, one needs 4 complex or 16 real multiplications. The
calculation of $\|H\|^2$ takes 4, its reciprocal $1/\| H\|^2$ 4, and
the multiplication of $1/\| H\|^2$ with ${\rm Re}\{{\rm Tr}[H^HA_k^HY]\}$ and
${\rm Im}\{{\rm Tr}[H^HB_k^HY]\}$ for $k=1,2$ another 4 real
multiplications. It can
be calculated that each of ${\rm Re}\{{\rm Tr}[H^HA_k^HY]\}$ and
${\rm Im}\{{\rm Tr}[H^HB_k^HY]\}$ has 3 distinct real additions for $k=1,2$,
which means there are a total of 12 real additions for this
operation. Calculation of $\|H\|^2$ takes 3 real additions. 
As a result, this approach employs 28 real multiplications and
15 real additions to decode. 

Note, in this case, the complexity figures in (\ref{eq-17}) 
are 28 real multiplications and 15 real additions, which hold exactly.
 
{\em Example 2:\/} Consider the OSTBC with $M=2$, $N=3$, $T=8$, and
$K=4$ given by \cite{tjc} 
\begin{equation}
{\cal G}_3=\left[
\begin{array}{cccccccc}
s_1&-s_2&-s_3&-s_4&s_1^*&-s_2^*&-s_3^*&-s_4^*\\
s_2&s_1&s_4&-s_3&s_2^*&s_1^*&s_4^*&-s_3^*\\
s_3&-s_4&s_1&s_2&s_3^*&-s_4^*&s_1^*&s_2^*\\
\end{array}
\right]^T .
\label{eq-28}
\end{equation}
The received signal can be written as 
\begin{equation}
\left[\begin{array}{cc}y_1^1 & y_1^2 \\ \vdots & \vdots \\
  y_8^1 & y_8^2\\\end{array}\right]=
{\cal G}_3
\left[\begin{array}{cc}h_{1,1} & h_{1,2}\\ h_{2,1} & h_{2,2} \\
h_{3,1} & h_{3,2} \\
\end{array}\right] +
\left[\begin{array}{cc} v_1^1 & v_1^2\\ \vdots & \vdots \\ 
  v_8^1 & v_8^2 \\
\end{array}\right] .
\label{eq-29}
\end{equation}
In \cite{aa2}, it has been shown that the $32\times 8$ real-valued
channel matrix $\check H$ is
\begin{equation}
\check H = \left [
\begin{array}{cccccccc}
h_1 & -h_2 & h_3 & -h_4 & h_5 & -h_6 & 0 & 0 \\
h_2 & h_1 & h_4 & h_3 & h_6 & h_5 & 0 & 0 \\
\vdots & \vdots & \vdots & \vdots & \vdots & \vdots & \vdots & \vdots\\
h_7& -h_8 &h_9 & -h_{10}&h_{11}&-h_{12}&0&0\\
h_8&h_7&h_{10}&h_9&h_{12}&h_{11}&0&0\\
\vdots & \vdots & \vdots & \vdots & \vdots & \vdots & \vdots & \vdots\\
0 & 0 & h_{11} & h_{12} & -h_9 & -h_{10} & -h_7 & -h_8\\
0 & 0 & h_{12} & -h_{11} & -h_{10} & h_9 & -h_8 & h_7 \\
\end{array}
\right ]
\label{eq-30}
\end{equation}
where $h_i$, $i=1,2,\ldots,11$ and $h_j$, $j=2,4,\ldots,12$ are the
real and imaginary parts, respectively, of $h_{1,1}$, $h_{2,1}$,
$h_{3,1}$, $h_{1,2}$, $h_{2,2}$, $h_{3,2}$. The matrix $\check H^T$ is
$8\times 32$ where each row has 8 zeros, while each of the remaining
24 symbols has one of $h_1,h_2,\ldots,h_{12}$, repeated twice. 
Let's first ignore the repetition of $h_i$ in a row. Then,
the calculation of $\check H^T\check y$ takes $8\cdot
24=192$ real multiplications.  
The calculation of $\sigma=\hat
h_1^T\hat h_1=2\sum_{k=1}^{12}h_i^2$ takes $12+1=13$ real multiplications, In
addition, one needs 
4 real multiplications to calculate $\sigma^{-1}$, and 8 real 
multiplications to 
calculate $\sigma^{-1}\bar y$. 
To calculate $\check H^T\check y$, one needs $8\cdot 23 = 184$ real
additions, and to calculate $\sigma$, one needs 11 real additions.
As a result, with this approach, one needs a total of 217 real multiplications
and 195 real additions to decode. 

For the method in (\ref{eq-20}) above, the products
$H^HA_1^H$ and $H^HB_1^H$ are 
\begin{equation}
\begin{array}{c}
H^HA_1^H = \left [
\begin{array}{cccccccc}
h_{1,1}^*&h_{2,1}^*&h_{3,1}^*&0&h_{1,1}^*&h_{2,1}^*&h_{3,1}^*&0\\
h_{1,2}^*&h_{2,2}^*&h_{3,2}^*&0&h_{1,2}^*&h_{2,2}^*&h_{3,2}^*&0\\
\end{array}
\right ] \\
H^HB_1^H=\left [
\begin{array}{cccccccc}
h_{1,1}^*&h_{2,1}^*&h_{3,1}^*&0&-h_{1,1}^*&-h_{2,1}^*&-h_{3,1}^*&0\\
h_{1,2}^*&h_{2,2}^*&h_{3,2}^*&0&-h_{1,2}^*&-h_{2,2}^*&-h_{3,2}^*&0\\
\end{array}
\right ]
\end{array}
\label{eq-31}
\end{equation}
Other $H^HA_k^H$ and $H^HB_k^H$ have similar structures, with zero
columns located elsewhere, same location in $H^HA_k^H$ and $H^HB_k^H$,
$k=2,3,4$. Nonzero columns of $H^HA_k^H$ and $H^HB_k^H$ are the
shuffled versions of the columns of $H^HA_1^H$ and $H^HB_1^H$, with
the same shuffling for $H^HA_k^H$ and $H^HB_k^H$, possibly with sign
changes. As a result, the first four columns of $H^HA_k^H$ and
$H^HB_k^H$ are the same, the first and second four columns of
$H^HA_k^H$ are the same, while the first and second four columns of
$H^HB_k^H$ are negatives of each other, $k=1,2,3,4$.
For this ${\cal G}_N$, one has 
\begin{equation}
{\cal G}_N^H{\cal G}_N=2\left ( \sum_{k=1}^K|s_k|^2\right ) I
\label{eq-32}
\end{equation}
which makes it necessary to replace $\| H\|^2$ with $2\| H\|^2$ in
(\ref{eq-20}) above.
The vector $Y$ is given as
\begin{equation}
Y=\left [ \begin{array}{cc}y_1^1&y_1^2\\\vdots&\vdots\\
y_8^1&y_8^2 \end{array} \right ] .
\label{eq-33}
\end{equation}
The complex multiplications in calculating 
${\rm Re}\{{\rm Tr}[H^HA_1^HY]\}$ can be used to calculate
${\rm Im}\{{\rm Tr}[H^HB_1^HY]\}$ due to sign changes and the calculation
of real and imaginary parts. Ignoring the repetition of $h_{i,j}^*$,
there are 12 different complex
numbers in $H^HA_1^H$ and due to the trace operation, they will be
multiplied with 12 complex numbers from $Y$. As a result, to calculate 
${\rm Tr}[H^HA_k^HY]$ (equivalently ${\rm Tr}[H^HB_k^HY]$)
one needs 12 complex or 48 real multiplications for one $k$. To
calculate the numerators of $s_k$, for all $k=1,2,3,4$, one needs 192 real
multiplications. To calculate $2 \| H\|^2$ in the denominator,
one needs 13 real multiplications. To calculate its inverse, one needs 4 real
multiplications. Finally, to complete the calculation of $s_k$ for
$k=1,2,3,4$ by 
multiplying the numerators of their real and imaginary parts by 
$1/(2 \| H\|^2)$, one needs 8 real multiplications. To calculate 
each ${\rm Re}\{{\rm Tr}[H^HA_k^HY]\}$ or ${\rm Im}\{{\rm Tr}[H^HB_k^HY]\}$ for
$k=1,2,3,4$, one needs $12 + 11 = 23$ real additions. To calculate
$\| H\|^2$, one needs 11 additions. 
As a result, with this approach, one needs 217 real multiplications
and 195 real additions to decode, same number as in the approach
specified by (\ref{eq-7}), (\ref{eq-10}), and (\ref{eq-15}).

For this example,
(\ref{eq-17}) specifies 300 real multiplications and 279 real additions. The
reduction is due to the elements with zero values in $\check H$.

It is important to make the observation that the repeated values of
$h_i$ in the columns of $\check H$, or equivalently $h_{m,n}^*$ in the
rows of $H^HA_k^H$ or $H^HB_k^H$, have a substantial impact on
complexity. We will carry out the rest of this discussion only for the
approach in (\ref{eq-7}), (\ref{eq-10}), and (\ref{eq-15}), the one in
(\ref{eq-20}) is essentially 
the same. Due to the repetition of $h_i$, by grouping the two values of
$\check y_j$ that it multiplies, it takes $8\cdot 12 = 96$ real
multiplications to compute $\check H^T\check y$, not $8 \cdot 24
=192$. The summations for each row of $\check H^T\check y$ will now be
done in two steps, first 12 pairs of additions per each $h_i$, and
then after multiplication by $h_i$, addition of 12 real numbers. This
takes $12 + 11 = 23$ real additions, with no change from the way the
calculation was made without grouping. With this change, the
complexity of decoding becomes 121 real multiplications and 195 real
additions, a huge reduction from 300 real multiplications and 279 real
additions. 

{\em Example 3:\/} We will now consider the code ${\cal G}_4$ from
\cite{tjc}. The parameters for this code are $N = K = 4$, $M=1$, and
$T=8$. It is given as
\begin{equation}
{\cal G}_4=\left[
\begin{array}{cccccccc}
s_1&-s_2&-s_3&-s_4&s_1^*&-s_2^*&-s_3^*&-s_4^*\\
s_2&s_1&s_4&-s_3&s_2^*&s_1^*&s_4^*&-s_3^*\\
s_3&-s_4&s_1&s_2&s_3^*&-s_4^*&s_1^*&s_2^*\\
s_4&s_3&-s_2&s_1&s_4^*&s_3^*&-s_2^*&s_1^*\\
\end{array}
\right]^T .
\label{eq-34}
\end{equation}
Similarly to ${\cal G}_3$ of Example 2, this code has the property
that ${\cal G}_4^H{\cal G}_4=2(\sum_{k=1}^K|s_k|^2)I$. As a result,
$\| H\|^2$ in the denominator in (\ref{eq-20}) should be replaced with
$2\| H\|^2$. The $\check H$ matrix is $16\times 8$ and can be
calculated as 
\begin{equation}
\check H = \left [
\begin{array}{cccccccc}
h_1 & -h_2 & h_3 & -h_4 & h_5 & -h_6 & h_7 & h_8 \\
h_2 & h_1 & h_4 & h_3 & h_6 & h_5 & h_8 & h_7 \\
h_3& -h_4 &-h_1 & h_{2}&h_{7}&-h_{8}&-h_5&h_6\\
h_4&h_3&-h_{2}&-h_1&h_{8}&h_{7}&-h_6&-h_5\\
\vdots & \vdots & \vdots & \vdots & \vdots & \vdots & \vdots & \vdots\\
h_5 & h_6 & -h_{7} & h_{8} & -h_1 & -h_{2} & h_3 & h_4\\
h_6 & -h_5 & -h_{8} & h_{7} & -h_{2} & h_1 & h_4 & -h_3 \\
\end{array}
\right ] .
\label{eq-35}
\end{equation}
This matrix consists entirely of nonzero entries. Each entry in a
column equals $\pm h_i$ for some $i\in \{1, 2,\ldots, 8\}$, every
$h_i$ appearing twice in a column. Ignoring this repetition for now,
calculation of $\check H^T\check y$ takes $8\cdot 16 = 128$ real
multiplications. Calculation of $\sigma$ takes 9 real multiplications, its
inverse 4 real multiplications, and the calculation of
$\sigma^{-1}\bar y$ takes 8 real multiplications. Calculation
of $\check H^T\check y$ takes $8\cdot 15=120$ real additions, and 
calculation of $\sigma$ takes 7 real additions. 
As a result, with this approach, to decode, one needs
149 real multiplications and 127 real additions.

For this code, for the method in (\ref{eq-20}), the
matrices $H^HA_k^H$ and $H^HB_k^H$ $k=1,2,3,4$ are as follows. 
\begin{eqnarray}
H^HA_1^H & = &
[\begin{array}{cccccccc}h_{1,1}^*&h_{2,1}^*&h_{3,1}^*&h_{4,1}^*&
      h_{1,1}^*&h_{2,1}^*&h_{3,1}^*&h_{4,1}^*\end{array}]
\nonumber\\
H^HA_2^H & = &
[\begin{array}{cccccccc}h_{2,1}^*&-h_{1,1}^*&-h_{4,1}^*&h_{3,1}^*&
      h_{2,1}^*&-h_{1,1}^*&-h_{4,1}^*&h_{3,1}^*\end{array}]
\nonumber\\
H^HA_3^H & = &
[\begin{array}{cccccccc}h_{3,1}^*&h_{4,1}^*&-h_{1,1}^*&-h_{2,1}^*&
      h_{3,1}^*&h_{4,1}^*&-h_{1,1}^*&-h_{2,1}^*\end{array}]
\nonumber\\
H^HA_4^H & = &
[\begin{array}{cccccccc}h_{4,1}^*&-h_{3,1}^*&h_{2,1}^*&-h_{1,1}^*&
      h_{4,1}^*&-h_{3,1}^*&h_{2,1}^*&-h_{1,1}^*\end{array}]
\label{eq-36}\\
H^HB_1^H & = &
[\begin{array}{cccccccc}h_{1,1}^*&h_{2,1}^*&h_{3,1}^*&h_{4,1}^*&
      -h_{1,1}^*&-h_{2,1}^*&-h_{3,1}^*&-h_{4,1}^*\end{array}]
\nonumber\\
H^HB_2^H & = &
[\begin{array}{cccccccc}h_{2,1}^*&-h_{1,1}^*&-h_{4,1}^*&h_{3,1}^*&
      -h_{2,1}^*&h_{1,1}^*&h_{4,1}^*&-h_{3,1}^*\end{array}]
\nonumber\\
H^HB_3^H & = &
[\begin{array}{cccccccc}h_{3,1}^*&h_{4,1}^*&-h_{1,1}^*&-h_{2,1}^*&
      -h_{3,1}^*&-h_{4,1}^*&h_{1,1}^*&h_{2,1}^*\end{array}]
\nonumber\\
H^HB_4^H & = &
[\begin{array}{cccccccc}h_{4,1}^*&-h_{3,1}^*&h_{2,1}^*&-h_{1,1}^*&
      -h_{4,1}^*&h_{3,1}^*&-h_{2,1}^*&h_{1,1}^*\end{array}]
\nonumber
\end{eqnarray}
From this set we conclude that the complex multiplications between
$H^HA_k^HY$ and $H^HB_kHY$ can be shared for a given $k=1,2,3,4$. The
number of real multiplications to calculate $H^HA_k^HY$ for all
$k=1,2,3,4$ is $4\cdot 8\cdot 4 = 128$. The number of real
multiplications to calculate $2 \| H\|^2$ is $6 + 1 = 7$, and to
calculate its inverse takes 4 real multiplications. Finally, the
number of real 
multiplications to complete the calculation of $s_k$ for all $k=1,2,3,4$ is
8. In order to calculate $H^HA_k^HY$ or 
$H^HB_k^HY$, one needs 8 real additions to perform each complex
multiplication and 7 real additions to calculate the sum. As a result, 
calculation of ${\rm Re}\{{\rm Tr}[H^HA_k^HY]\}$ and 
${\rm Im}\{{\rm Tr}[H^HB_k^HY]\}$ for all $k=1,2,3,4$ takes $8\times 15$
real additions. Calculation of $\| H\|^2$ takes
$6+1=7$ real additions. 
Therefore, with this approach the number of real multiplications
and additions to decode are 149 and 127, respectively, same as the
numbers needed for the approach in (\ref{eq-7}), (\ref{eq-10}), and
(\ref{eq-15}).  

For this example, equation (\ref{eq-17}) specifies 156 real
multiplications and 135 real additions. The reduction is due to the
fact that one row of $\check H^T$ has each $h_i$ appearing twice. This
reduces the number of multiplications and summations to calculate
$\sigma$ by about a factor of 2.

However, because each $h_i$ appears twice in every row of $\check H^T$, the
number of multiplications can actually be reduced substantially, as we
discussed in Example 2. As discussed in Example 2, we can reduce the
number of multiplications to calculate $\check H^T\check y$ by
grouping the two multipliers of each $h_i$ by summing them prior to
multiplication by $h_i$, $i=1,2,\ldots,8$. As seen in Example 2, 
this does not alter the number of real additions. With this simple
change, the number of real multiplications to decode becomes 85 and
the number of real additions to decode remains at 127.

{\em Example 4:\/} It is instructive to consider the code ${\cal H}_3$
given in \cite{tjc} with $N=3,$ $K=3$, $T=4$ which we will consider
for $M=1$ where
\begin{equation}
{\cal H}_3=\left[
\begin{array}{ccc}
s_1&s_2&{s_3}/{\sqrt{2}}\\
-s_2^*&s_1^*&{s_3}/{\sqrt{2}}\\
{s_3^*}/{\sqrt{2}}&{s_3^*}/{\sqrt{2}}&(-s_1-s_1^*+s_2-s_2^*)/{2}\\
{s_3^*}/{\sqrt{2}}&{-s_3^*}/{\sqrt{2}}&(s_2+s_2^*+s_1-s_1^*)/{2}\\
\end{array}
\right].
\label{eq-37}
\end{equation}
For this code, ${\cal H}_3^H{\cal H}_3=(\sum_{k=1}^3|s_k|^2)I$ is satisfied.
In this case, the matrix $\check H$ can be calculated as
\begin{equation}
\check H=\left[
\begin{array}{cccccc}
h_1&-h_2&h_3&-h_4&h_5/\sqrt{2}&-h_6/\sqrt{2}\\
h_2&h_1&h_4&h_3&h_6/\sqrt{2}&h_5/\sqrt{2}\\
h_3&h_4&-h_1&-h_2&h_5/\sqrt{2}&-h_6/\sqrt{2}\\
h_4&-h_3&-h_2&h_1&h_6/\sqrt{2}&h_5/\sqrt{2}\\
-h_5&0&0&-h_6&(h_1+h_3)/\sqrt{2}&(h_2+h_4)/\sqrt{2}\\
-h_6&0&0&h_5&(h_2+h_4)/\sqrt{2}&-(h_1+h_3)/\sqrt{2}\\
0&h_6&h_5&0&(h_1-h_3)/\sqrt{2}&(h_2-h_4)/\sqrt{2}\\
0&-h_5&h_6&0&(h_2-h_4)/\sqrt{2}&(-h_1+h_3)/\sqrt{2}\\
\end{array}
\right].
\label{eq-38}
\end{equation}
It can be verified that every column $\check h_i$ of $\check H$ has
the property that $\check h_i^T\check
h_i=\sigma=\|H\|^2=\sum_{k=1}^6h_k^6$ for $i=1,2,\ldots,6$. In this
case, the number of real multiplications to calculate $\check H^T\check y$
requires more caution than the previous examples. For the first
four rows of $\check H^T$, this number is 6 real multiplications per
row. For the last two rows, due to combining, e.g., $h_1$ and $h_3$ in 
$(h_1+h_3)/\sqrt{2}$ in the fifth element of $\check h_5$, and the
commonality of $h_5$ and $h_6$ for the first and third, and second and
fourth, respectively, elements of $\check h_5$, and one single multiplier
$1/\sqrt{2}$ for the whole column, the number of real multiplications
needed is 7. As a result, calculation of $\check H^T\check y$ takes 38
real multiplications. Calculation of $\sigma$ takes 6 real 
multiplications. One needs 4 real multiplications to calculate $\sigma^{-1}$,
and 6 real multiplications to calculate $\sigma^{-1}\bar y$. First
four rows of $\check 
H^T\check y$ require 5 real additions each. Last two rows of $\check
H^T\check y$ require $4+7=11$ real additions each. This is a total of
42 real additions to calculate $\check H^T\check y$. Calculation of
$\sigma$ requires 5 real additions. Overall, with this approach one
needs 54 real multiplications and 47 real additions to decode. 

For this code, for the method in (\ref{eq-20}) above, the
matrices $H^HA_k^H$ and $H^HB_k^H$, $k=1,2,3$ are as follows. 
\begin{eqnarray}
H^HA_1^H & = &
[\begin{array}{cccc}h_{1,1}^*&h_{2,1}^*&-h_{3,1}^*&0\end{array}]\nonumber\\
H^HA_2^H & = &
[\begin{array}{cccc}h_{2,1}^*&-h_{1,1}^*&0&h_{3,1}^*\end{array}]\nonumber\\ 
H^HA_3^H & = & \frac{1}{\sqrt{2}}
[\begin{array}{cccc}h_{3,1}^*&h_{3,1}^*&h_{1,1}^*+h_{2,1}^*&h_{1,1}^*-h_{2,1}^*\end{array}] \\\label{eq-39}
H^HB_1^H&=&[\begin{array}{cccc}h_{1,1}^*&-h_{2,1}^*&0&h_{3,1}^*\end{array}]\nonumber\\
H^HB_2^H&=&[\begin{array}{cccc}h_{2,1}^*&h_{1,1}^*&h_{3,1}^*&0\end{array}]\nonumber\\
H^HB_3^H&=&\frac{1}{\sqrt{2}}
[\begin{array}{cccc}h_{3,1}^*&h_{3,1}^*&-h_{1,1}^*-h_{2,1}^*&-h_{1,1}^*+h_{2,1}^*\end{array}]\nonumber
\end{eqnarray}

Before discussing the complexity of the approach in (\ref{eq-20}), we
would like to make an observation. 
A careful examination shows that the complex multiplications between 
$H^HA_k^HY$ for $k=1,2,3$ and $H^HB_j^HY$ for $j=1,2,3$ can be shared in
the method outlined in (\ref{eq-20}). In this case, since
$h_3^*$ in the first and second element of $H^HA_3^H$ can be shared, there
are 9 complex multiplications needed for the calculation of 
$H^HA_k^HY$ for $k=1,2,3$. The real values of those will be used in
calculating the real parts of $s_k$, $k=1,2,3$, and the imaginary
parts in calculating the imaginary parts of $s_k$, $k=1,2,3$, albeit
in possibly different signs or locations. This requires a careful
implementation where the needed complex multiplications are
calculated, stored, and their real and imaginary parts carefully
distributed in the most judicious manner. The 9 complex
multiplications correspond to 36 real multiplications, and there are
2 more real multiplications, by $1/\sqrt{2}$ for the real and
imaginary parts of $s_3$. As in the previous method, 6 real
multiplications are needed to calculate $\|H\|^2$, 4 real
multiplications to calculate $1/\|H\|^2$, and 6 real multiplications
to complete the calculation of $s_k$, $k=1,2,3$. 
The calculation of
${\rm Re}\{{\rm Tr}[H^HA_k^HY]\}$ and ${\rm Im}\{{\rm Tr}[H^HB_k^HY]\}$ for
all $k=1,2,3$ takes $4\cdot 5 + 2 \cdot (6+5) = 42$ real additions, 
and the calculation of $\| H\|^2$ takes 5 more real additions. 
This approach results in a total of 54 real multiplications and 47
real additions to decode, as in the technique in 
(\ref{eq-7}), (\ref{eq-10}), and (\ref{eq-15}). 

For this example, (\ref{eq-17}) specifies 66 real multiplications and 49
real additions. The reduction is due to the presence of the zero
entries in $\check H$. On the other hand, the presence of the factor
$1/\sqrt{2}$ in the last two rows of $\check H^T$ adds two real
multiplications to the total number of real multiplications.

Before concluding this example, we would like to display the matrices
$A_3$ and $B_3$ for this code. 
\begin{equation}
A_3=\left[\begin{array}{ccc}
0&0&1/\sqrt{2}\\
0&0&1/\sqrt{2}\\
1/\sqrt{2}&1/\sqrt{2}&0\\
1/\sqrt{2}&-1/\sqrt{2}&0\\
\end{array}\right] 
\hspace{1cm}
B_3=\left[\begin{array}{ccc}
0&0&1/\sqrt{2}\\
0&0&1/\sqrt{2}\\
-1/\sqrt{2}&-1/\sqrt{2}&0\\
-1/\sqrt{2}&1/\sqrt{2}&0\\
\end{array}\right] 
\label{eq-40}
\end{equation}
In all other $A_k$ and $B_k$ matrices in the four examples studied,
the entries were $\pm 1$. Furthermore, in all other $A_k$ and $B_k$
matrices in the four examples, there was at most one nonzero value in
a row. In both $A_3$ and $B_3$ above, the entries are irrational 
numbers and two rows have two nonzero entries.

From the examples above, by studying the operations of the two
techniques in detail, it can actually be seen that, not only is the
computational complexity of the technique in (\ref{eq-7}),
(\ref{eq-11}), and (\ref{eq-15}) is the same as the technique in
(\ref{eq-20}), but also they actually perform equivalent
operations. 
\section{Orthogonality of $\check H$ and Computational Complexity
  Revisited}
We have seen in the examples that when 
${\cal G}_N^H{\cal G}_N=c(\sum_{k=1}^K |s_k|^2)I$ where $c=1,2$, then 
$\sigma=c\|H\|^2$. We will now show this holds in general. Based on
that result, we will then reduce the computational complexity estimate in
(\ref{eq-17}). 

Let 
\begin{equation}
z={\rm vec}(Y)= \left[\begin{array}{c}y_1^1\\\vdots\\ y_T^M \end{array}\right].
\label{eq-41}
\end{equation}
Form two vectors, $\bar s$ and $\tilde s$,
consisting of real and imaginary parts of $s_k$, and
form a vector $s'$ that is the concatenation of $\bar s$ and $\tilde s$:
\begin{equation}
\bar s = (\bar s_1 , \bar s_2 , \ldots , \bar s_K)^T,\hspace{4mm}
\tilde s = (\tilde s_1 , \tilde s_2 , \ldots , \tilde s_K)^T,\hspace{4mm}
s'=(\bar s , \tilde s)^T.
\label{eq-42}
\end{equation}
By rearranging the right hand side of (\ref{eq-2}), we can write
\begin{equation}
z = F s' + e = F_a \bar s + F_b \tilde s + e
\label{eq-43}
\end{equation}
where $F=[F_a \hspace{1mm}F_b]$ is an $MT\times 2K$, and $F_a$ and $F_b$ are
$MT\times K$ complex matrices whose entries consist of (linear
combinations of) channel coefficients $h_{i,j}$, and $e$ is the
corresponding complex Gaussian noise vector. In \cite{ls}, it was
shown that when ${\cal G}_N^H{\cal G}_N=(\sum_{k=1}^K |s_k|^2)I$, then
${\rm Re} [ F^HF]=\| H\|^2 I$. It is straightforward to extend this result so
that when ${\cal G}_N^H{\cal G}_N=c (\sum_{k=1}^K |s_k|^2)I$, then 
\begin{equation}
{\rm Re} [ F^HF]=c \| H\|^2 I
\label{eq-44}
\end{equation}
where $c$ is a positive integer. Let 
\begin{equation}
\bar z = {\rm Re}[z], \hspace{4mm}\tilde z = {\rm Im}[z],\hspace{4mm}
\bar e = {\rm Re}[e], \hspace{4mm}\tilde e = {\rm Im}[e],
\label{eq-45}
\end{equation}
and
\begin{equation}
\bar F_a={\rm Re}[F_a],\hspace{4mm}\tilde F_a={\rm Im}[F_a],\hspace{4mm}
\bar F_b={\rm Re}[F_b],\hspace{4mm}\tilde F_b={\rm Im}[F_b].
\label{eq-46}
\end{equation}
Now define
\begin{equation}
z'=\left[\begin{array}{c}\bar z\\\tilde z \end{array}\right]
\hspace{4mm}
F'=\left[\begin{array}{cc}\bar F_a & \bar F_b\\\tilde F_a &\tilde F_b
\end{array}\right]\hspace{4mm}
e'=\left[\begin{array}{c}\bar e\\\tilde e\end{array}\right]
\hspace{4mm}
\label{eq-47}
\end{equation}
so that we can write
\begin{equation}
y' = F' s' + e'
\label{eq-48}
\end{equation}
which is actually the same expression as (\ref{eq-4}) except
the vectors and matrices have their rows and columns permuted. 

It can
be shown that (\ref{eq-44}) implies
\begin{equation}
F'\hspace{.5mm}^TF'=c\|H\|^2 I.
\label{eq-49}
\end{equation}

Let $P_y$ and $P_s$ be $2K\times 2K$ and $2MT\times 2MT$,
respectively, permutation matrices such that 
\begin{equation}
\left[\begin{array}{c}{\rm Re}(y_1^1)\\{\rm Im}(y_1^1) \\ \vdots \\
 {\rm Re}(y_T^M)\\ {\rm Im}(y_T^M) \end{array} \right] = P_yy'\hspace{4mm}
\left[\begin{array}{c}{\rm Re}(s_1)\\{\rm Im}(s_1) \\ \vdots \\
 {\rm Re}(s_K)\\ {\rm Im}(s_K) \end{array} \right] =P_ss' . 
\label{eq-50}
\end{equation}
It follows that $P_y^TP_y^{\ }=P_y^{\ }P_y^T=I$ and 
$P_s^TP_s^{\ }=P_s^{\ }P_s^T=I$.

We now have
\begin{eqnarray}
\left[\begin{array}{c}{\rm Re}(y_1^1)\\{\rm Im}(y_1^1) \\ \vdots \\
 {\rm Re}(y_T^M)\\ {\rm Im}(y_T^M) \end{array} \right]& = & 
P_y(F's'+e')\\
& = & P_yF'P_s^T
\left[\begin{array}{c}{\rm Re}(s_1)\\{\rm Im}(s_1) \\ \vdots \\
 {\rm Re}(s_K)\\ {\rm Im}(s_K) \end{array} \right] +P_ye' \\
& = &
\check H
\left[\begin{array}{c}{\rm Re}(s_1)\\{\rm Im}(s_1) \\ \vdots \\
 {\rm Re}(s_K)\\ {\rm Im}(s_K) \end{array} \right] 
+
\left[\begin{array}{c}{\rm Re}(v_1^1)\\{\rm Im}(v_1^1) \\ \vdots \\
 {\rm Re}(v_T^M)\\ {\rm Im}(v_T^M) \end{array} \right]  .
\label{eq-53}
\end{eqnarray}
Therefore,
\begin{equation}
\check H = P_yF'P_s^T
\label{eq-54}
\end{equation}
which implies
\begin{equation}
\check H^T\check H = P_s^{\ }F'\hspace{0.5mm}^TP_y^TP_y^{\ }F'P_s^T
=c \| H \|^2 I.
\label{eq-55}
\end{equation}
In other words, $\sigma=c\| H\|^2$.
This has an impact on the computational complexity formula
(\ref{eq-17}) which we discuss next.

First, let $c=1$. Since $\sigma=\| H\|^2$, its calculation takes $2MN$ 
real multiplications and $2MN-1$ real additions. As a result, the
computational complexity formula (\ref{eq-17}) can be updated as
\begin{equation}
{\cal C}_{\rm PR}=(4KMT+2MN+2K+4)R_M,(4KMT+2MN-2K-1)R_A .
\label{eq-56}
\end{equation}

When $c>1$, the number of real multiplications to calculate $\sigma$
increases by 1, however, the complexity of the calculation of $\check
H^T\check y$ will reduce by a factor of $c$, as seen in the
examples. 

As seen in the examples, the presence of values of 0
within $\check H$ will reduce the computational 
complexity. Its effect will be a reduction in the number of real
multiplications to calculate $\check H^T\check y$ by a factor equal to
the ratio of 
the rows of $A_k$ and $B_k$ that 
consist only of 0 values to the total number of all rows in $A_k$ and
$B_k$ for $k=1,2\ldots,K$, with a similar (not same) reduction in the
number of real additions to calculate $\check H^T\check y$. It will
also reduce the number of real multiplications and additions to
calculate $\sigma$ but that effect can be more complicated, as seen in
Example 4. Also, as seen in Example 4, the contents of the $\check H$
matrix can have linear combinations of $h_i$ values, which also result
in changes in 
computational complexity. 
\section{Discussion}
For an OSTBC ${\cal G}_N$ satisfying 
${\cal G}_N^H{\cal G}=c(\sum_{k=1}^K|s_k\|^2)I$ where $c$ is a
positive integer, the Maximum Likelihood solution is formulated in
four equivalent ways 
\begin{equation}
\|Y-{\cal G}_NH\|^2=\|z-Fs'\|^2=\|z'-F's'\|^2=\|\check y-\check Hx\|^2.
\label{eq-57}
\end{equation}
There are four solutions, all equal. The first solution is obtained by
expanding $\|Y-{\cal G}_NH\|^2$ and is given by (\ref{eq-20}) when
$c=1$ [4, eq. (7.4.2)]. When $c>1$, it should be altered as
\begin{equation}
\hat s_k = \frac{1}{c \| H\|^2}[{\rm Re}\{{\rm Tr}(H^HA_k^HY)\}-\hat\imath\cdot
  {\rm Im}\{{\rm Tr}(H^HB_k^HY)\}] \hspace{1cm}k=1,2,\ldots,K.
\label{eq-58}
\end{equation}
The second solution is obtained by expanding the second expression in
(\ref{eq-57}) and is given by 
\begin{equation}
\hat s'=\frac{{\rm Re}[F^Hz]}{c\| H\|^2} .
\label{eq-59}
\end{equation}
This is given in [4. eq. (7.4.20)] for $c=1$. The third solution is
the solution to the third equation in (\ref{eq-57})
\begin{equation}
\hat s'=\frac{F'\hspace{.5mm}^T z'}{c \|H\|^2} .
\label{eq-60}
\end{equation}
The fourth solution is the one introduced in \cite{aa}. It is the
solution to the fourth equation in (\ref{eq-57}) and is given by
\begin{equation}
\left[\begin{array}{c}{\rm Re}(\hat s_1)\\{\rm Im}(\hat s_1) \\ \vdots \\
 {\rm Re}(\hat s_K)\\ {\rm Im}(\hat s_K) \end{array} \right] 
=\frac{\check H^T \check y}{\sigma}
=\frac{\check H^T \check y}{c\| H\|^2} .
\label{eq-61}
\end{equation}
Considering that
\begin{equation}
F_a=[{\rm vec}(HA_1)\hspace{1mm}\cdots \hspace{1mm}{\rm vec}(HA_K)]\hspace{4mm}
F_b=[\hat\imath {\rm vec}(HB_1)\hspace{1mm}\cdots
  \hspace{1mm}\hat\imath {\rm vec}(HB_K)] 
\end{equation}
[4, eq. (7.1.7)], it can be verified that (\ref{eq-58}) and 
(\ref{eq-59}) are equal. The equality of (\ref{eq-59}) and
(\ref{eq-60}) follows from (\ref{eq-45})-(\ref{eq-47}). The equality of
(\ref{eq-60}) and (\ref{eq-61}) follows from (\ref{eq-50}) and (\ref{eq-54}).
Therefore, equations (\ref{eq-58})-(\ref{eq-61}) yield the same
result, and when 
properly implemented, will have identical computational complexity.

Finally, we would like to state that a straightforward implementation of
(\ref{eq-58}) or (\ref{eq-59}) 
can actually result in larger complexity than (\ref{eq-60}) and
(\ref{eq-61}). The proper implementation requires that in
(\ref{eq-58}) 
and (\ref{eq-59}), the terms not needed due to elimination by the Tr[ ],
Re[ ], and Im[ ] operators are not calculated. We calculated the
computational complexity values for the examples taking this fact into account.
\bibliographystyle{IEEEtran}
\bibliography{IEEEabrv,bib/Ayanoglu} 

\end{document}